\documentclass[12pt]{article}

\usepackage{amsmath}
\usepackage{amssymb}

\begin{document}

\title{Change in the Nuclear Moment of Inertia at the Number of Neutrons $N = 98$ and
Isotopic Shift of Atomic Levels
\footnote{Yad. Fiz. {\bf 61} (1), 17--120 (1998) [Phys. Atomic Nuclei,
{\bf 61,} No.~1, 13--15 (1998)]}}

\author{ A.M. Kamchatnov$^{\dagger }$and V.G. Nosov$^{\ddagger}$\\
$^{\dagger}${\small\it Institute of Spectroscopy, Russian Academy of Sciences,
Troitsk, Moscow Region, 142190 Russia}\\
$^{\ddagger}${\small\it Russian Research Center Kurchatov Institute, pl. Kurchatova 1,
Moscow, 123182 Russia}
}

\maketitle

\begin{abstract}
From analysis of experimental data on nuclear rotation, it is found that the
so-called rigid-body value of the nuclear moment of inertia and the nuclear-radius
values associated with the moment of inertia as functions of the number of nucleons
change their behavior at the number of neutrons $N = 98$. This phenomenon is
confirmed by experimental data on the isotopic shifts of atomic levels.
\end{abstract}

It is well known that the nuclear radius $R$ as a function of the number $A$ of
nucleons is described by the simple relation
\begin{equation}
\label{eq(1)}
R=r_0 A^{1/3},
\end{equation}
which has a clear geometric interpretation: the nuclear volume is proportional
to the number of nucleons. For the majority of nuclei the constant $r_0$ takes the value of
\begin{equation}
\label{eq(2)}
r_0 = 1.1\cdot 10^{-13} \mbox{cm},
\end{equation}
but it can slightly depend on the method of its experimental determination.
In particular, it is the value in (2) that is obtained from analysis of experimental
data on electron scattering on nuclei and from isotopic shifts of atomic levels [1].

It is, however, difficult to verify relation (1) by analyzing the dynamical
properties of nuclei, although some relevant features of the nucleus---for example,
the nuclear moment of inertia---obviously depend on the nuclear radius. Shortly after
the discovery of rotational bands in nuclei, it was found that the effective moment of
inertia $\mathcal{J}$ which can be determined from the relation
\begin{equation}
\label{eq(3)}
\frac{\hbar ^2}{{\cal J}(I)} = \frac{d^2 E(I)}{dI^2},
\end{equation}
where $E(I)$ is the energy of the rotational level with angular momentum $I$,
differs considerably from the rigid-body value
\begin{equation}
\label{eq(4)}
{\cal J}_0 =\frac{2}{5}MR^2,
\end{equation}
where $M$ is the nuclear mass, while $R$ is the nuclear radius (2). For example,
at the head of a rotational band,
the effective moment of inertia (3) is usually two or three times as small as the
corresponding rigid-body value [2]. It was found in [3] that, with increasing
angular momentum $I$, the moment of inertia $\mathcal{J}$ also increases and may become
several times as great as the value in (4) near the backbending point. This result was
confirmed later by numerous experimental and theoretical studies (for an overview,
see [4-6]). By analyzing available experimental data, we found [7,8] that the
aforementioned quantities obey simple semi-empirical relations involving the rigid-body
value of the moment of inertia. This makes it possible to verify relation (1) in
dynamical phenomena.

In the simplest way, the aforementioned relations are expressed in terms of the behavior
of the nuclear angular velocity
\begin{equation}
\label{eq(5)}
\hbar\Omega =\frac{dE(I)}{dI}
\end{equation}
as a function of the angular momentum $I$ with respect to the so-called
rigid-body line
\begin{equation}
\label{eq(6)}
\hbar\Omega =\frac{\hbar^2}{{\cal J}_0} I.
\end{equation}
Namely, experimental data demonstrate that the backbending on the curve representing
the dependence of the angular velocity $\Omega$ on the angular momentum $I$
is always accompanied by the intersection of the rigid-body line in a downward
direction. Therefore, above the backbending point $I_c$, the angular velocity lies
below the corresponding rigid-body values (6). At the backbending point $I_c$,
the moment of inertia (3) often exhibits a singular behavior represented by the function
\begin{equation}
\label{eq(7)}
{\cal J}\simeq\frac{j}{I-I_c}, \qquad I\to I_c+0.
\end{equation}

Above the backbending point, the angular velocity takes the rigid-body value
\begin{equation}
\label{eq(8)}
\hbar\Omega_{nc} = \frac{\hbar^2}{{\cal J}_0} I_c .
\end{equation}
As $I$ increases further moving away from $I_c$, we come to the asymptotic region
where the moment of inertia is close to the rigid-body value
\begin{equation}
\label{eq(9)}
{\cal J}\simeq {\cal J}_0, \qquad I-I_c \gg j,
\end{equation}
and the angular velocity approaches the rigid-body line from below. Such asymptotic
behavior is violated by so-called repeated backbendings. Equations (8) and (9) make
it possible to calculate the rigid-body value of the moment of inertia from
experimental data to a rather high accuracy and to determine thereby the nuclear
radius $R$ from the dynamical properties of nuclear rotation. In this way, the value
of $r_0$ in (2) was confirmed once again for the majority of nuclei.

Nevertheless, it was demonstrated in [9] that the rigid-body value of the moment of
inertia decreases sharply at the point where the number of neutrons is $N = 98$.
As a result, the parameter $r_0$ takes the value
\begin{equation}
\label{eq(10)}
r_0=1.0\cdot 10^{-13}\mbox{cm}.
\end{equation}
for nuclei in which the number of neutrons varies from $N = 98$ to the magic number
$N = 126$, which corresponds to the transition to spherical nuclei. Experimental data
obtained later confirm this conclusion. Figure 1 shows the nuclear angular velocity
as a function of the angular momentum for the yrast lines of the
tungsten isotopes (a) $^{166}_{74}$W$_{92}$ [10], (b) $^{168}_{74}$W$_{94}$ [11],
(c) $^{170}_{74}$W$_{96}$ [12], (d) $^{172}_{74}$W$_{98}$ [13],
(e) $^{174}_{74}$W$_{100}$ [14], and (f) $^{179}_{74}$W$_{105}$ [15].
It can easily be seen that the character of experimental data changes for the isotopes
in which the number of neutrons is greater than $N = 98$. At the backbending point,
the curve representing $\hbar\Omega$ does not intersect the rigid-body line (6) for
the moment of inertia corresponding to the value in (2). However, the aforementioned
regularities are fulfilled for $N\geq 98$ as well if the rigid-body lines correspond
to the radius value in (10). Similar changes in the moment of inertia at $N = 98$ occur
for the rotational bands of Hf and Os isotopes. We note that this phenomenon is
observed not only for yrast-lines but also for rotational bands built on excited
nuclear states. This confirms our assumption [7--9] that the first backbending points
are of a universal nature that is independent of the properties of the rotational
band under consideration.

This brings about the natural question of whether it is possible to verify the
aforementioned changes in the nuclear radius at $N = 98$ by some other, independent,
methods. The nuclear shape directly affects isotopic shifts of atomic levels.
Unfortunately, the isotopic shifts in W, Hf, and Os isotopes have not yet been
measured to a sufficient accuracy because of experimental difficulties. However,
there are comprehensive data on the isotopic shifts in Yb isotopes [16]. Figure 2
shows the mean-square nuclear radius as a function of the number $N$ of neutrons.
This curve has break at $N = 98$, and the radius increases with $N$ more slowly for
$N\geq 98$ than for $N \leq 98$. This corresponds to a decrease in $r_0$ for $N \geq 98$.
Of course, the correlation between the two phenomena may be only qualitative because
the isotopic shifts depend on the distribution of the electric charge within a nucleus,
whereas the moments of inertia are determined by the mass distribution (these two
distributions obviously do not coincide). Nonetheless, the above data indicate that
the phenomena being discussed are correlated.

\section*{Acknowledgements}

We are grateful to A.V.~Zybin and A.V.~Mishin for stimulating discussions of
experimental data on isotopic shifts of atomic levels.

\bigskip

\bigskip

\centerline{\bf Figures captions}

\bigskip

Fig.~1. Angular velocities for the yrast lines of $W$ isotopes as functions of the
orbital angular momentum $I$. The corresponding $r_0$ values are indicated on the
rigid-body lines.

\bigskip

Fig. 2. Deviation of the mean-square nuclear radius of Yb isotopes from the value
for the nucleus with $N = 98$ versus the number of neutrons according to measured
isotopic shifts of atomic levels.

\end{document}